\documentclass[11pt]{article}
\usepackage{a4wide}
\usepackage{amsmath}
\usepackage{amssymb}
\usepackage{natbib}
\usepackage{color}
\usepackage{bm}
\usepackage{graphics,graphicx}
\usepackage{url}
\usepackage{hyperref}
\usepackage{pdfpages}
\usepackage{multirow}
\makeatletter

\newcommand{\para}{\bigskip\noindent}

\newcommand{\balpha}{\bm{\alpha}}
\newcommand{\bbeta}{\bm{\beta}}
\newcommand{\beps}{\bm{\epsilon}}

\newcommand{\btheta}{\bm{\theta}}

\newcommand{\blambda}{\bm{\lambda}}

\newcommand{\by}{\mathbf{y}}

\newcommand{\be}{\begin{equation}}
\newcommand{\ee}{\end{equation}}
\newcommand{\TITLE}[1]{{\center{\bf#1}\vskip30pt}}

\newcommand{\AFFILIATION}[1]{{\small\center {\it #1} \\}}
\newcommand{\AUTHORS}[1] {{\small\center{#1}\\}}

\begin{document}

\TITLE{Guiding adaptive shrinkage by co-data to improve regression-based prediction and feature selection
}

\AUTHORS{Mark A. van de Wiel$^1$, Wessel N. van Wieringen$^{1,2}$}

\AFFILIATION{$^1$Dept of Epidemiology and Data Science, Amsterdam Public Health research
institute, Amsterdam University Medical Centers, Amsterdam, The Netherlands;
$^2$Dept of Mathematics, VU University, Amsterdam, The Netherlands
}

\begin{abstract}
The high dimensional nature of genomics data complicates feature selection, in particular in low sample size studies - not uncommon in clinical prediction settings.
It is widely recognized that complementary data on the features, `co-data', may improve results. Examples are prior feature groups or p-values from a related study.
Such co-data are ubiquitous in genomics settings due to the availability of public repositories. Yet, the uptake of learning methods that structurally use such co-data is limited. We review guided adaptive shrinkage methods: a class of regression-based learners that use co-data to adapt the shrinkage parameters, crucial for the performance of those learners. We discuss technical aspects, but also the applicability in terms of types of co-data that can be handled. This class of methods is contrasted with several others. In particular, group-adaptive shrinkage is compared with the better-known sparse group-lasso by evaluating feature selection. Finally, we demonstrate the versatility of the guided shrinkage methodology by showing how to `do-it-yourself': we integrate implementations of a co-data learner and the spike-and-slab prior for the purpose of improving feature selection in genetics studies.
\end{abstract}

\section{Introduction}
Genetics and genomics data are usually of a high-dimensional nature: the number of measured features vastly exceeds the number of samples.
Two plagues of such high-dimensional data are low signal-to-noise ratio and multicollinearity. All prediction and feature selection models are affected by those plagues. Here, we focus on regression-based models, which employ regularization by introducing shrinkage either through a penalty or an informative prior. Let us first elaborate on these two plagues and their implications before discussing potential cures.

\para
First, a low signal-to-noise ratio implies an abundance of irrelevant features. As, by default, shrinkage parameters are shared by all features, such an abundance may lead to overshrinkage for relevant features. On its turn, this may harm the predictive accuracy of the resulting model.
The second one is multicollinearity, which doubles in omics as many genomic features are highly correlated due to shared biological properties. In a predictive model features are competing with one another. Selecting one feature will likely de-select another one if the two are strongly correlated. Then, small fluctuations in the data set drive the feature selection, rendering it instable.

\para
Three cures for the two plagues come immediately to mind. First, the low signal-to-noise ration is countered by enforcing sparsity, e.g. by a lasso penalty or a horseshoe prior, to better accommodate a strong contrast between relevant and non-relevant features. This certainly helps when there indeed exists such a strong contrast, but may be less appropriate in many genomics settings in which many small effects may pile up \cite[]{boyle2017expanded}. That is, in the latter case one may wish to accommodate the grey scale between relevant and non-relevant. Second, the instable feature selection due to multicollinearity may be countered by stability selection \cite[]{meinshausen2010stability}. In a nutshell, one generates many random copies of the data set, e.g. by bootstrapping, and then composes the ultimate set of selected features from those which are selected in a large proportion of those copies. The third solution is to bring in external knowledge. The promises of this solution are clear: external information allows for better modelling of the shrinkage and can break the strong competition between features. This solution may be combined with the former ones.

\para
We argue that the last solution is still under-used in practice, which is why we focus on it in this review. One reason for the under-use is pragmatism: it takes time to compile external data and requires thinking on what to include. Moreover, leveraging external knowledge can be achieved in many different ways, rendering it difficult to have an overview and pick an algorithm for one's needs. Our aim is two-fold: on one hand convince potential users that leveraging external knowledge may be well worth the effort and on the other hand provide guidance on which algorithms to use for which settings.

\para
Our focus lies on methods that allow guided adaptive shrinkage. That is, the shrinkage is modeled as a function of the external information on the features. We refer to the latter as `co-data' \cite[]{Neuenschwander2010, WielGRridge}, `complementary data'. Alternatively, the term `features of features' has been coined \cite[]{tay2023feature}.
The group-adaptive lasso \cite[]{zeng2021incorporating, van2023fast} is a special case of guided adaptive shrinkage. As the (sparse) group-lasso is a better-known method that applies to the same setting, i.e. one co-data source defining feature groups, we start by contrasting the penalty functions of these two methods. Then, we shift the focus to a general formulation of guided adaptive shrinkage, and review several methods. In particular, we consider what types of co-data can be handled, what types of response are accommodated, and what strategy is used to estimate hyperparameters.

\para
We contrast the guided adaptive shrinkage approach to related ones. These either share the adaptive nature of the former - such as the adaptive lasso - or the use of co-data, such as structured regularization methods (including the sparse group-lasso), and regression-based transfer learning. As the contrast between group-adaptive lasso and group-lasso is particularly relevant, we use simulations to compare the two for a varying number of feature groups in terms of feature selection performance. Throughout, we focus on evaluating feature selection, as the potential benefit of using co-data for the purpose of prediction has already been demonstrated by many of the discussed works.

\para
Finally, we show the versatility of the approach by discussing a `do-it-yourself' solution for one's favorite model. We use it to illustrate the benefit of co-data for feature selection in a spike-and-slab model, which is a popular Bayesian model for selecting features in large scale genomics studies \cite[]{carbonetto2012scalable}.

%


\section{Group-adaptive lasso versus sparse group lasso}
\begin{figure}[h]
\begin{center}
\includegraphics[scale=0.5]{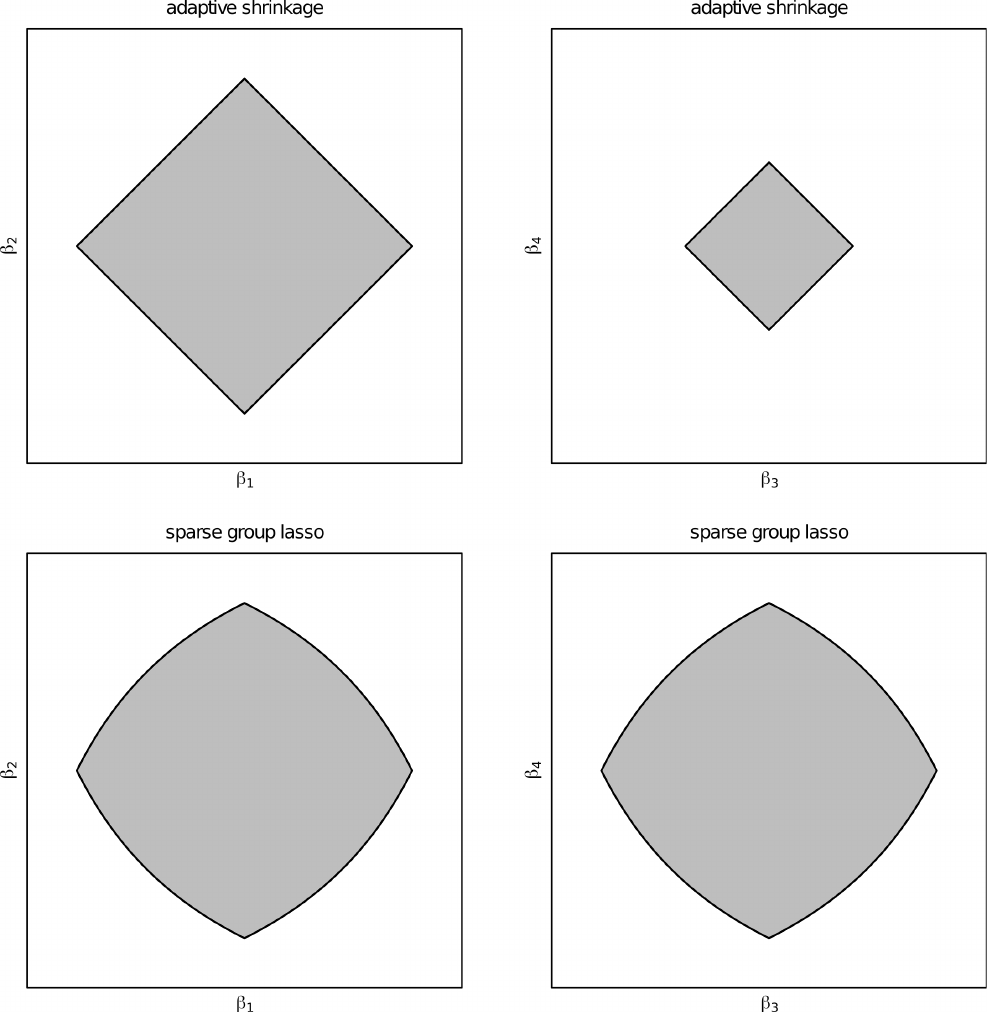}
\end{center}
\caption{Parameter constraints for two groups of two features. Top-row: group-adaptive lasso; bottom-row: the sparse group lasso}\label{constraints}
\end{figure}

Before defining the guided adaptive shrinkage methodology in a general framework, we briefly discuss a canonical setting: co-data that consists of one grouping
of the features. Examples are different data modalities - all used within one regression - with possibly very different dimensions (e.g. gene expression, mutations, clinical variables, imaging-derived features) or a grouping based on genomic location, such as the chromosomes.  Both the group-adaptive lasso \cite[]{zeng2021incorporating, van2023fast} and sparse group lasso \cite[]{simon2013sparse} can accommodate such groups, denoted by $G_g, g=1, \ldots, G$. The essential difference between the two is adaptivity.
Sparse group-lasso extends the lasso by augmenting the L$_1$ penalty function on the regression coefficients of the $p$ features, $(\beta_j)_{j=1}^p$, with a group-penalty on the $G$ groups of features, whereas
group-adaptive lasso employs different penalties across groups of features. The penalty functions $P$ for group-lasso and its adaptive counterpart are:
\begin{align}
P_{\lambda, \lambda'}(\bbeta) &= \lambda \sum_{j=1}^p |\beta_j| + \lambda' \sum_{g=1}^G ||\bbeta_g||_2 & & \text{(sparse group-lasso)} \label{sgl}\\
P_{\blambda}(\bbeta) &= \sum_{g=1}^G \lambda_g \sum_{j \in G_g} |\beta_j| & & \text{(group-adaptive lasso)} \label{gal},
\end{align}
with norm $||\bbeta_g||_2 = (\sum_{k \in G_g} \beta_{k}^2)^{1/2}$.
Penalized regression can also be cast in an equivalent constraint optimization setting, where the constraints are one-to-one linked to the penalties.
Figure \ref{constraints} illustrates these constraints for a toy example with two groups of two features. Clearly, the constraint adapts to the strength of a group of features in the group-adaptive setting, whereas the constraint is essentially the same for both groups in the sparse group-lasso setting. Those constraints are very important in high-dimensional settings, because features compete to be selected. Hence, depending on the situation, the two methods may perform very differently, as we will illustrate further on.

\section{Guided adaptive shrinkage}
Here, we introduce and illustrate the general framework before reviewing a variety of guided adaptive shrinkage methods.
\subsection{General framework}
\begin{figure}[h]
\begin{center}
\includegraphics[scale=0.9]{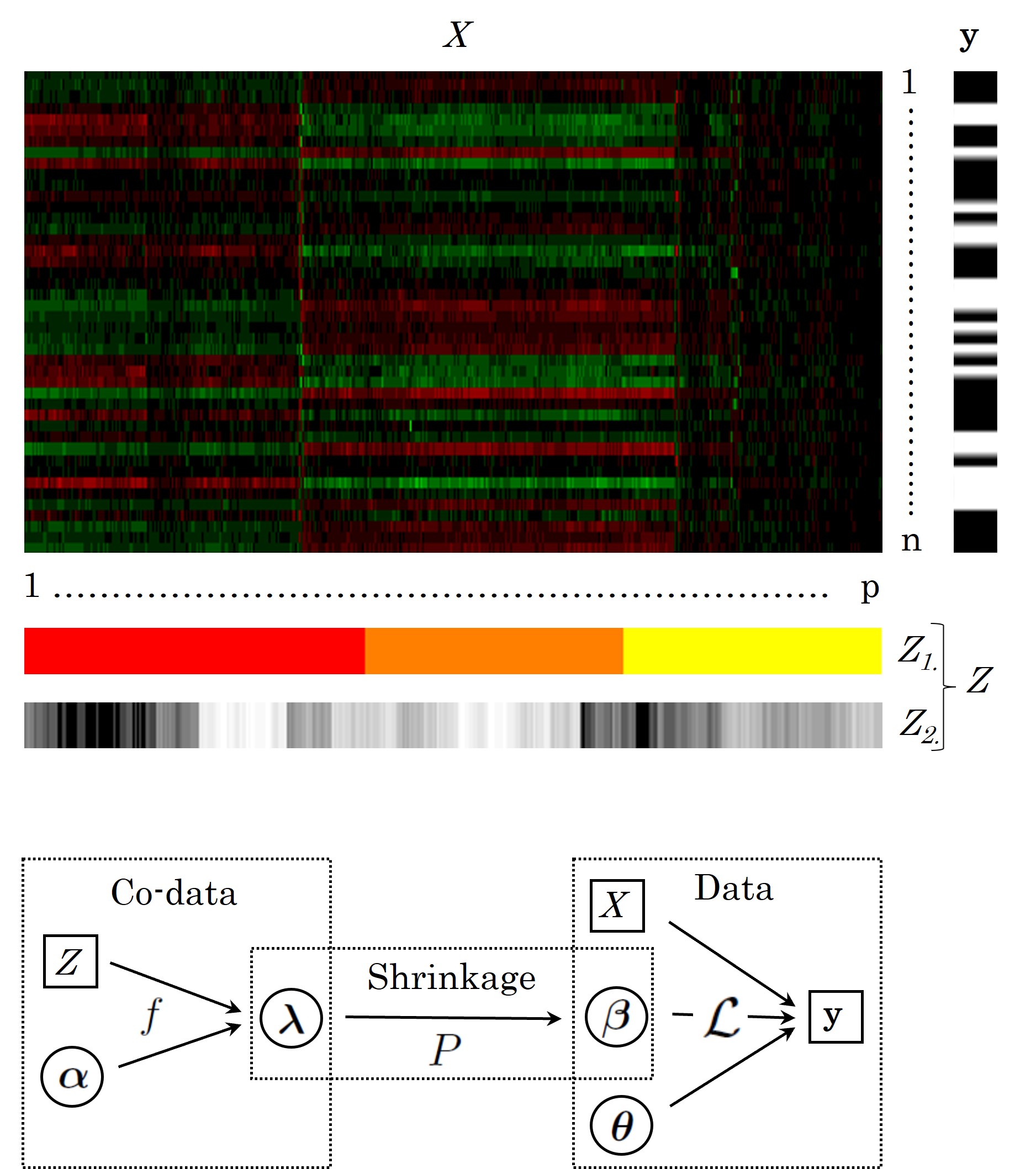}
\end{center}
\caption{Guided adaptive shrinkage using co-data $Z = \big(\begin{smallmatrix}
  Z_{1.}\\
  Z_{2.}
\end{smallmatrix}\big)$. $\balpha$: hyper-parameters; $f$: adaptation function; $\blambda$: penalty vector; $P$: shrinkage via penalty or prior; $\bbeta$: regression parameters;  $X$: data; $\btheta$: nuisance parameters; $\mathcal{L}$: likelihood; $\by$: response}\label{codatillustr}
\end{figure}

Let $\by = (y_i)_{i=1}^n$ be the response of interest, $X = (x_{ij})_{i=1,j=1}^{n,p}$ the data matrix with $x_{ij}:$ the value of feature $j$ for sample $i$, $\bbeta = (\beta_j)_{j=1}^p$ the regression coefficients, and $\btheta = (\theta_1, \ldots, \theta_k)$ denote the nuisance parameters (such as noise variance $\sigma^2$). We assume that the regression model linking $X$ to $\by$ by $\bbeta$ defines a likelihood function $\mathcal{L}$. Other fit functions could be used as well. Then, the guided adaptive shrinkage framework is summarized by:

\begin{align*}
&\mathcal{L}(\bbeta, \btheta; X,\by) & &\text{(Likelihood)}\\
&P_{\blambda}(\bbeta) & &\text{(Shrinkage)}\\
&\lambda_j = f_{\balpha}(Z_{.j}) & &\text{(Guided adaptation)},
\end{align*}
with penalty vector $\blambda = (\lambda_1, \ldots, \lambda_p)$, and co-data matrix $Z = (z_{cj})_{c,j=1}^{C,p}$. $Z$ comprises of rows $Z_{c.}$ that corresponds to co-data source $c$ and of columns $Z_{.j}$ corresponding to feature $j$. $P_{\blambda}(\bbeta)$ is either a penalty function in a classical setting, or a prior in a Bayesian setting. In fact, when the penalty function is formulated as a log-prior, maximization of the penalized likelihood renders the Bayesian posterior mode estimate. This equivalence facilitates switching between the two paradigms. Besides the choice of paradigm and type of shrinkage, methods differ in terms of how they estimate $\blambda$ and how they incorporate co-data matrix or vector $Z$. Here, co-data function $f$, parameterized by a lower dimensional parameter $\balpha$, connects $Z$ to the penalties $\blambda$. We emphasize the relative low dimension of $\balpha$ (w.r.t. $p$): this facilitates stable estimation of the high-dimensional penalty vector $\blambda$. Figure \ref{codatillustr} illustrates guided adaptive shrinkage.

\newpage


\subsection{Methods}
Guided adaptive shrinkage methods can be classified in multiple ways. The likelihood $\mathcal{L}$, largely determined by the type of response $\by$ (e.g. continuous/binary/survival), and the type of penalty/prior $P_{\blambda}$ are obvious classifiers. Moreover, the type of co-data $Z$ allowed by co-data function $f_{\balpha}$ differentiates the applicability of methods in two ways: can multiple co-data sources be accommodated, and what types are allowed: grouped, continuous or mixed? Finally, the methods crucially differ in how hyperparameters are estimated. Here, we distinguish 1) Cross-validation: hyperparameters are tuned by cross-validation; 2) Empirical Bayes: hyperparameters are tuned by empirical Bayes techniques; 3) Full Bayes: hyperparameters are endowed with an hierarchical prior; 4) Joint estimation: Hyperparameters and regression parameters are estimated jointly. Table \ref{cotabel} classifies co-data guided shrinkage methods according to these criteria. Below we provide more specific details on each of these methods.

\begin{table}[ht]
\begin{tabular}{lllllll}
Method & Reference$^{*}$ & Software & Likelihood$^{\small{+}}$ & Shrinkage & Co-data & Hyp.$^{\$}$ \\\hline\hline
Multi- & Tai & \texttt{-} & Bin & Lasso & Group & CV\\
pen. PLS & (\citeyear{Tai2007}) & & & & Uni & \\\hline
Weighted & Bergensen  & \texttt{-} & Gauss, & Lasso & Cont & CV\\
lasso & (\citeyear{Bergersen2011}) & & Bin, Cox & & Uni & \\\hline
Group-regul. & vd Wiel  & \texttt{GRridge} & Gauss & Ridge & Group & EB\\
ridge & (\citeyear{WielGRridge}) & & Bin, Cox & & Multi & \\\hline
Integrated pen. & Boulesteix  & \texttt{ipflasso} & Gauss & Lasso & Group & CV\\
factors lasso & (\citeyear{boulesteix2017ipf}) & & Bin, Cox & & Uni &\\\hline
Group-adapt. & Velten  & \texttt{graper} &  Gauss & Lasso, Ridge & Group & FB\\
pen. regr. & (\citeyear{velten2018adaptive}) & & Bin & Spike \& Slab & Uni & \\\hline
Group-regul. & Ignatiadis  & \texttt{SigmaRidge} & Gauss & Ridge & Group & EB+\\
ridge & (\citeyear{ignatiadis2020sigma}) & (Julia) & & & Uni &  CV\\\hline
Co-data adapt. & Van Nee  & \texttt{ecpc} & Gauss & Ridge & Mixed & EB\\
ridge & (\citeyear{van2021flexible}) & & Bin, Cox & & Multi & \\\hline
Incorp. prior & Zeng  & \texttt{xtune} & Gauss & Elastic Net & Mixed & EB\\
info pen. regr.& (\citeyear{zeng2021incorporating}) & & Bin, Mult & & Multi &\\\hline
Hierarchical & Kawaguchi & \texttt{xrnet} & Gauss & Ridge & Mixed & CV\\
ridge & (\citeyear{kawaguchi2022hierarchical}) & & GLM & &  Multi & \\\hline
Group-adapt. & Van Nee & \texttt{squeezy} & Gauss & Elastic Net & Group & EB\\
elastic net & (\citeyear{van2023fast}) & & Bin, Cox & & Uni & \\\hline
Feature-weight. & Tay  & \texttt{fwelnet}  & Gauss & Elastic net & Mixed & Joined \\
elastic net & (\citeyear{tay2023feature}) & & Bin & & Multi & \\\hline
Co-data adapt. & Busatto & \texttt{infHS} & Gauss & Horseshoe & Mixed & FB\\
Horseshoe regr. & (\citeyear{busatto2023informative}) & & Probit & & Multi & \\\hline
\end{tabular}
$^{*}\,:$ First author only\\
$^{\small{+}}:$ Gauss: Gaussian, continuous; Bin: Binomial, binary; Cox: Proportional hazards, survival \\
$^{\$}\,:$ Hyperparameter estimation. CV: Cross-validation; EB: Empricial Bayes; FB: Full Bayes \\
\caption{Co-data adaptive shrinkage methods}\label{cotabel}
\end{table}

\bigskip
\texttt{Multi-penalty PLS} \cite[]{Tai2007} is a pioneering method on the adaptation of penalties based on grouped co-data.
It models co-data function $f_{\balpha}$ simply by $\balpha = (\lambda_1, \ldots, \lambda_G)$. The authors cast their method in a partial least squares setting, but they
show the correspondence to ordinary regression. The method uses soft-thresholding with adaptive thresholds, which is strongly related to group-adaptive lasso regression. It accommodates only one grouped co-data source. It introduces an heuristic to reduce computing time for tuning $\blambda$ by cross-validation, using a weighting function.

 \texttt{Weighted lasso} \cite[]{Bergersen2011} models the co-data function $f_{\balpha}= f_{\lambda, q}$, which depends on \`a priori defined similarity weights between $X$ and $Z$ or between $\by$ and $Z$. The weighting function is somewhat arbitrary and limited in flexibility. It relies on two tuning parameters: the lasso penalty $\lambda$ and $q$, which tunes the importance of the weights. Two examples of similarity weights are: correlation between an mRNA feature ($X$) and its DNA counterpart ($Z$), or regression coefficients that relate $\by$ to $Z$. This method is simple, easy to extend and implement. Moreover, it is relatively fast, because only two hyper-parameters are tuned. It accommodates only one co-data source, though, that needs to be continuous.

 \texttt{GRridge} and \texttt{ecpc} \cite[]{WielGRridge,van2021flexible} are both based on ridge regression, but the latter extends on the former by allowing non-grouped co-data using a regression parametrization, $f_{\balpha} = Z_{.j}\balpha$). In addition, \texttt{ecpc} implements hyperparameter shrinkage, which is useful when a co-data source consists of many feature groups. The methods share the methodology for hyperparameter estimation by moment-based empirical Bayes. The estimation procedure is modular, which provides computational efficiency and flexibility in terms of implementation, but does not propagate uncertainty as full Bayesian procedures do. The setting is not sparse, although posterior variable selection is implemented, and shown to be competitive to lasso-based methods.

 \texttt{ipflasso} \cite[]{boulesteix2017ipf} is a group-adaptive method, so $f_{\balpha}$ is simply parameterized by $\balpha = (\lambda_1, \ldots, \lambda_G)$. The method is simple and easy to extend. As it tunes $\blambda$ by full-blown multi-grid cross-validation it may be slow when the number of feature groups increases. It accommodates only one grouped co-data source. The methods is extended by \cite{zhao2020structured} to allow for hierarchical groups.

 \texttt{graper} \cite[]{velten2018adaptive} is the first fully Bayesian co-data method that is flexible in terms of penalization and incorporates, next to lasso and ridge priors, the spike-and-slab. It is less flexible in terms of co-data as it only allows groups. Computational scalability is achieved by developing a variational Bayes approximation. As a fully Bayesian method it provides uncertainty quantification, although this was not evaluated by the authors and may be compromised by the use of variational Bayes.

\texttt{SigmaRidge} \cite[]{ignatiadis2020sigma} is a hybrid method that deals with group-adaptive shrinkage in the ridge setting. It combines cross-validation and empirical Bayes for hyperparameter tuning. In the Bayesian formulation, the error noise in the linear model, $\sigma$, is also a hyperparameter. Here, it is treated as a global parameter tuned by CV, whereas $\blambda = (\lambda_1, \ldots, \lambda_G)$ is determined by moment-based empirical Bayes. As the latter is analytical, it is embedded in the cross-validation of $\sigma$. This leads to truly joint estimation of all hyperparameters. Asymptotic optimality results are provided and computational efficiency is achieved by approximate, analytical leave-one-out-CV. The scope of the method is limited, though, as it only handles linear regression with grouped co-data.

\texttt{xtune} \cite[]{zeng2021incorporating} supports the use of generic, mixed-type co-data by modeling $f_{\balpha} = Z_{.j}\balpha$. It provides efficient estimation of hyperparameters $\balpha$ by a Gaussian approximation of the elastic net prior. Moreover, it uses a majorization technique to speed up optimizing of the marginal likelihood to tune $\balpha$ by empirical Bayes. Hence, it is computationally efficient and versatile in its use. Results are presented for linear response only, but the software supports binary and multi-class response as well.

\texttt{xrnet} \cite[]{kawaguchi2022hierarchical} differs from the other ones in the way it shrinks $\bbeta$: to a co-data moderated \emph{mean}
$Z_{.j} \balpha$ instead of a variance. It shrinks $\balpha$ to zero within the same objective function using a hierarchical ridge penalty. It is computationally efficient, as once the two hyperparameters are fixed, the objective function is convex. It does not provide a solution for feature selection.

\texttt{squeezy} \cite[]{van2023fast} is similar in spirit to \texttt{xtune}, as it also approximates the marginal likelihood by a Gaussian one. It compliments this approximation, however, with a proof based on the multivariate central limit theorem. It is computationally very efficient as it makes use of many shortcuts available in the Gaussian setting. The main implementation supports grouped co-data only, but it can directly use the output of ridge-based \texttt{ecpc} (discussed above) to handle mixed co-data and hyperparameter shrinkage.

\texttt{fwelnet} \cite[]{tay2023feature} models, conditionally on a global penalty parameter $\lambda$, feature-specific penalties by $f_{\balpha}= \lambda w_j(\balpha),$
with weights $w_j(\balpha) = \bigl(\exp(Z_{.j} \balpha) / \sum_{j=1}^p \exp(Z_{.j} \balpha)\bigr)^{-1}$. The approach uniquely optimizes hyperparameters $\balpha$ and regression parameters $\bbeta$ jointly. The optimization algorithm alternates between updating $\balpha$ and $\bbeta$, using gradient search for $\balpha$ and an elastic net solver for $\bbeta$. A potential identification problem is circumvented by normalizing the weights. Hence, it is a non-hierarchical formulation, which adapts weights instead of shrinkage.


\texttt{infHS} \cite[]{busatto2023informative} is a fully Bayesian method that uses the popular horseshoe prior to encode sparsity into the predictive model. It uses a regression parametrization ($Z_{.j}\balpha$) to allow mixed co-data types. It modifies the prior mean of the local regularization parameters, thereby particularly facilitating high-dimensional settings with many small signals and a few outlying large ones. It is suitable for feature selection, and the variational Bayes approximation of the posteriors provides computational scalability of the method. Binary outcome is accommodated by a probit formulation.


\section{Related methods}
Below we discuss some related regression-based methods that either share the use of external knowledge or the concept of adaptation of shrinkage with the discussed guided adaptive shrinkage methodology.
\subsection{Group penalties, structured regularization}
Guided adaptive shrinkage relates to structured regularization as both frameworks allow to incorporate external information in the regularization of regression models. Well-known examples of the latter are the group-lasso and the hierarchical lasso, but many more methods are available; see \cite{vinga2021structured} for a extensive review, and \cite{zhu2019bayesian} for a Bayesian perspective. As both the sparse group-lasso (and variations thereof) and the group-adaptive lasso can be applied to co-data groups, we restrict ourselves to a comparison between those two types of methods with penalty functions \eqref{sgl} and \eqref{gal}. Here, co-data matrix $Z$ consists of only one row vector $Z_{1.}$ containing categorical entries that correspond to the feature groups.

The sparse group-lasso focuses on selecting groups and features, whereas the group-adaptive lasso selects only features while adapting to different group strengths. Hence, the former may be more suitable in settings with many groups, of which a large part is not relevant, while the latter is more flexible in settings with few groups. As an illustration, we briefly study this claim in a simulation setting.

In a linear regression setting, $\by = X\bbeta + \beps$, we simulate $n=200$ samples for $p=2,000$ features divided over $G=3, 6, 9, 15, 24, 39, 60, 99$ equally-sized groups. Of these groups, 1/3rd contains non-zero coefficients. To allow some variation between the non-zero groups, the proportion of non-zero coefficients in those groups is sampled from a Beta(2,6) distribution, averaging to 1/4 non-zero's per group. Non-zero $\beta$'s are generated from a scaled $t_3$ distribution, such that the total explained variation of the features equals that of the Gaussian noise, $\epsilon_i \sim N(0,\sigma^2=1)$. Features $x_{ij}$ are independently sampled from a standard Gaussian.

\begin{figure}[h]
\begin{center}
\includegraphics[scale=0.44]{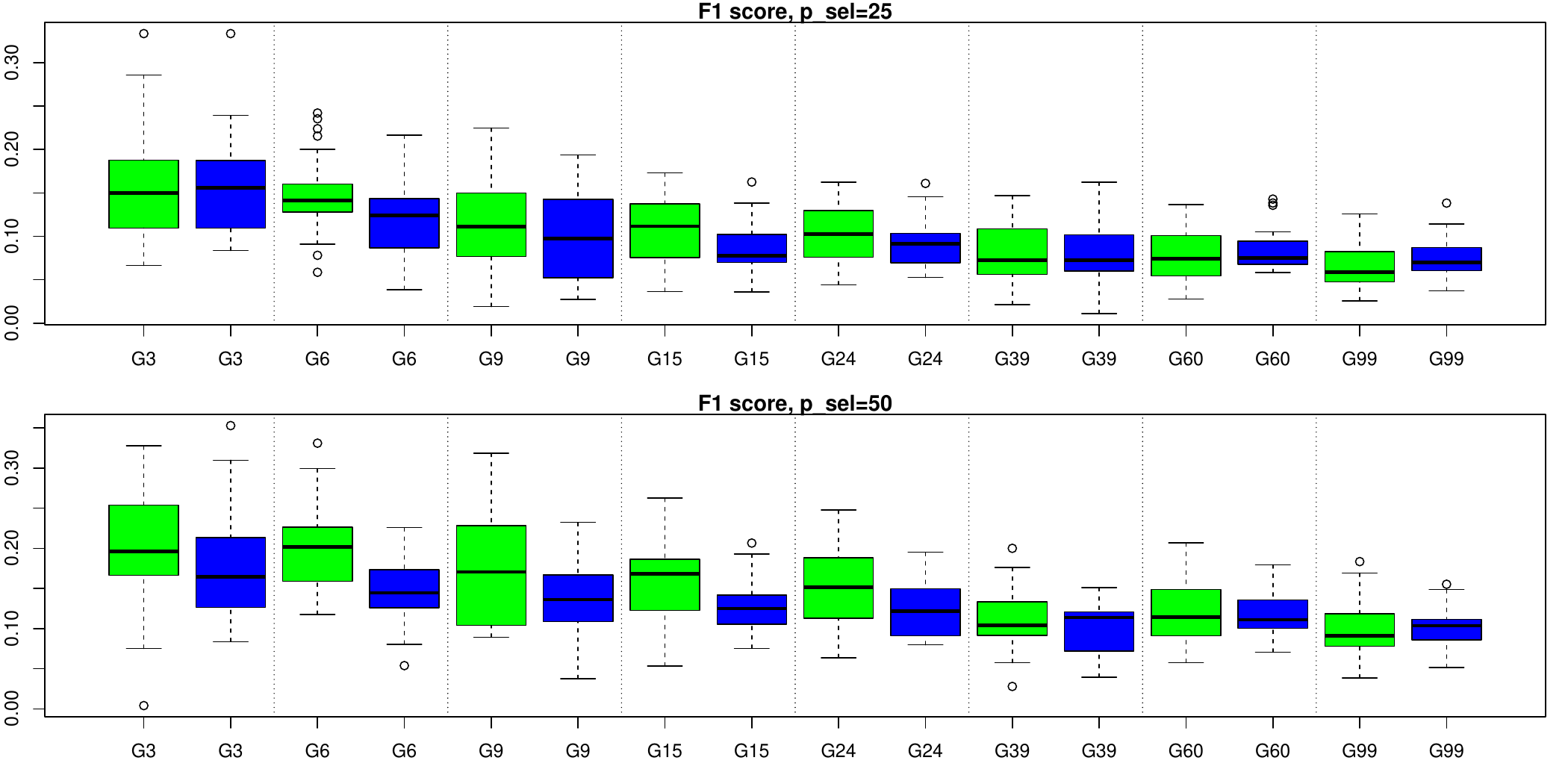}
\end{center}
\caption{F1 score for variable selection ($p_{\text{sel}} = 25, 50$) for group-adaptive lasso (green) and sparse group-lasso (blue) for $G =3, \ldots, 99$ feature groups (25 simulations per $G$)}\label{F1}
\end{figure}

On the simulated training data, the group-adaptive lasso and the sparse group-lasso were fitted using the R packages \texttt{squeezy} and \texttt{SGL}, respectively, using the known feature groups as input and using defaults for other parameters. We focus on feature selection by evaluating the F1-score, the harmonic mean of precision and recall.
As such scores are somewhat incomparable for models of different size, we opt to fix the number of selected non-zero features, $p_{\text{sel}}$. Both methods allow this as both produce a regularization path that may be used for this purpose. For each group-size $G$, simulations were repeated 25 times. Figure \ref{F1} shows the results for $p_{\text{sel}} = 25, 50$.

The simulations clearly support the claim: for a small to intermediate number of groups, group-adaptive lasso outperforms sparse group-lasso, whereas the latter becomes superior when many groups are used. Within one setting, group-adaptive lasso is generally somewhat more variable in performance across repeats, possibly due to the higher number of hyperparameters that need to be estimated.

The Supplementary Material shows an extra simulation scenario that is more `group-sparse': $p = 10,000, G= 60, 99$ and 5 groups with non-zero coefficients. These results support our conclusion above: for a large number of groups ($G=99$) sparse group-lasso is somewhat superior to group-adaptive lasso, but the latter is competitive for an intermediate number of groups ($G=60$), even in this fairly group-sparse scenario. Finally, the Supplementary Material also provides a solution to shrink hyperparameters in the group-adaptive lasso setting. This is shown to be particularly useful when the number of groups is large and when these groups are not informative (the `null-setting').

\subsection{Adaptive lasso and variations}
A second class of methods related to guided adaptive shrinkage, consists of the adaptive lasso \cite[]{Zou2006}, and variations thereof. There is a fundamental difference between the two though. The former uses external data to guide the adaptation, whereas the latter bases the adaptation purely on the data itself: it uses the parameter estimates of an unpenalized (or L$_2$-penalized) model as inverse penalty weights. The adaptive lasso has proven its use in low-to-medium dimensional settings, in which it can counter the overpenalization of non-zero parameters by the lasso. In high-dimensional settings, however, double use of the same data may come at the cost of overfitting. The extend of the latter likely depends on the (unknown) underlying sparsity. \cite{belhechmi2020accounting} present a hybrid adaptive/group-adaptive algorithm which may (partly) overcome the overfitting issue. Basically, it groups the parameters estimates of an initial fit according to the prior groups, and uses these to define weights in the adaptive step.

\subsection{Regression-based transfer learning}
A third class of alternative methods fits under the umbrella of regression-based transfer learning. Transfer learning is a large subfield of machine learning that aims to transfer knowledge from one domain to another. Without the aim to be complete we discuss three methods in the regression framework that apply to high-dimensional data; we refer to those papers for further reading. These three methods all transfer previously found regression coefficients \emph{directly} to the regression coefficients at hand. Some of the discussed guided adaptive shrinkage methods also accommodate such continuous co-data (external regression coefficients), but these methods transfer this information \emph{indirectly}: to the shrinkage instead of to the coefficients.

\para
First, \cite{boonstra2013incorporating} discuss targeted ridge shrinkage: $\bbeta$ is shrunken to a non-zero mean $\bbeta_0$, which are regression coefficients obtained from a similar prediction problem. Hence, somewhat similar in spirit as in \cite[]{kawaguchi2022hierarchical}, but the latter can use more general co-data to modify the target.
Second, the prior lasso \cite[]{jiang2016variable} transfers predictions from a prior model into the objective function: it augments the likelihood with a weighted likelihood part that uses the predictions from the prior model as the outcome. This renders one estimate $\hat{\bbeta}$ that accommodates both the primary data and, to a lesser extent, the prior data, the impact of which can be tuned.
Third, \texttt{transreg} \cite[]{rauschenberger2023penalized} estimates regression coefficients as a function of the prior coefficients available from co-data. It uses either a specific parameteric form, or a non-parametric one that respects monotony to ensure stability. It bags the prior-informed predictors, possibly from multiple co-data sources, with a prior-agnostic one to create a meta-learner, and optimizes the weights by efficient cross-validation.

\subsection{Other related methods}
Finally, we discuss a few other methods that are related to guided adaptive shrinkage. First, priority-lasso \cite[]{klau2018priority} also handles multiple groups of features (e.g. data modalities), but it prioritizes some groups over others, a strategy also explored by \cite{aben2016tandem}. The idea is that a given data modality may have some practical advantages over other ones: e.g. (stable) DNA markers may be preferred over less stable mRNA markers.
The Integrated Elastic Net \cite[]{culos2020integration} is used in the context of immune response prediction and is similar in spirit as \texttt{fwelnet}, although it optimizes the hyperparameters \emph{separately} from $\bbeta$ using cross-validation. It focuses on tensored binary co-data that codes for cell types, stimulations and type of response.
In \cite{yang2023tsplasso}, external information is also used to improve feature selection, but only after initial feature selection by lasso. It uses the prior information to determine a set of features that is relatively stable and relatively well in line with that information.
Finally, \cite[]{aldahmani2020graphical} present a graphical-group ridge. It uses a graphical model to determine network modules. Its nodes represent feature groups in a ridge regression with group-specific penalties.

\section{Do-it-yourself for your favorite model}
The guided adaptive shrinkage papers reviewed in Table \ref{cotabel} focus on the most popular penalties/priors, in particular variations of the elastic net. Many other penalties and priors have been proposed, which triggers the question whether these can easily be modified to allow for adaptive shrinkage. For many methods, this is indeed the case. For MCMC-based methods, one generic solution is extensively discussed in \cite{WielEB}, building upon an algorithm in \cite{Casella2001empirical}. They show that the hyperparameters which link the co-data to the priors can be estimated by alternating MCMC with likelihood-based optimization. The method is conceptually straightforward, but can be very time-consuming, as it requires multiple MCMC runs. Below, we focus on a `do-it-yourself' solution that is computationally much more efficient.

\para
Our solution applies when the penalty corresponds to a prior with finite variance and a non-sparse data matrix $X$. Gene expression data usually satisfies the latter condition, whereas data on very rare mutations may not. \cite{van2023fast} prove that under these conditions a Gaussian approximation of the prior is asymptotically appropriate to estimate the hyperparameters $\balpha$. This corresponds to ridge regression for which very efficient algorithms are available to determine the hyperparameters by maximizing the marginal likelihood. This method has been implemented for the elastic net in the R-packages \texttt{xtune}\cite[]{zeng2021incorporating} and \texttt{squeezy}\cite[]{van2023fast}. It can, however, also be used to estimate other penalties/priors using the following algorithm:

\begin{enumerate}
  \item Determine co-data sources $Z$ and define Gaussian (= ridge) variances $v^R_j = \lambda_j^{-1}$, with $f(\lambda_j) = Z_{.j}\balpha$
  \item Estimate $\balpha$, and hence $v^R_j$, in the Gaussian setting using \texttt{xtune} or \texttt{squeezy}
  \item Equate the theoretical variance $v_j$ of the desired prior to $\hat{v}^R_j$ to obtain feature-specific prior parameters
  \item Estimated the high-dimensional model using those feature-specific prior parameters
\end{enumerate}

\para
Next, we give an example for the spike-and-slab prior.
This prior is a versatile, natural and powerful prior for high-dimensional settings, in particular useful for Bayesian feature selection \cite[]{carbonetto2012scalable, newcombe2014weibull, velten2018adaptive}. We illustrate the `do-it-yourself' principle in this setting: the co-data guided spike-and-slab prior.
For this purpose, we focus on the simplest formulation of the spike-and-slab prior:
\begin{equation}\label{ssprior}
\beta_j \sim (1-q) \delta_0 + q N(0,\tau^2),
\end{equation}
with an appropriate prior on the global parameter $\tau^2$. \cite{carbonetto2012scalable} derive a computationally very efficient variational Bayes algorithm to approximate posteriors of $\bbeta$, and use its implementation, \texttt{varbvs}, for feature selection in very high-dimensional genetic studies.
Now suppose one wishes to incorporate co-data sources $Z_{1.} = (z_{1j})_{j=1}^{p}$ and $Z_{2.} = (z_{2j})_{j=1}^{p}$  to modify the prior inclusion probability, $q$. E.g. $Z_{1.}$ presents a prior grouping of SNPs into two groups, while $Z_{2.}$ represents log-p-values for those SNPs in a previous study. Then, we may modify the prior to:
\begin{equation}\label{ssprnew}
\beta_j \sim (1-q_{j}) \delta_0 + q_{j} N(0,\tau^2),
\end{equation}
where the feature specific inclusion probability depends on the co-data for feature $j$: $Z_{.j} = (z_{1j}, z_{2j})$. This prior is available in \texttt{varbvs}, but requires specification of $(q_j)_{j=1}^p$. We now explain how to estimate this quantity with the existing
software tools following the algorithm above.

\para
First, we model the ridge penalties $\lambda_j = f_{\balpha}(Z_{.j}) = \alpha_1 z_{1j} + \alpha_2 z_{2j}$. Second, we estimate $(\alpha_1, \alpha_2)$ by using the R-package \texttt{xtune}. The reciprocal penalties render ridge-based variances $\hat{v}^R_j$. Third, we compute the theoretical prior variances $v_j$ from \eqref{ssprnew}. For that, denote the latent indicator $I_{\{\beta_j=0\}}$ by $I_j$. Then, we have for the prior variance of $\beta_j$:
$$v_j = V(\beta_j) = E_{I_j}[V[\beta_j | I_j]] + V_{I_j}[E[\beta_j | I_j]] = (1-q_j)*0 + q_j\tau^2 + 0 = q_j\tau^2.$$
Equating $v_j$ to $\hat{v}^R_j$ renders relative estimates of $q_j$, as we have $q_j = C v_j$, with $C$ an unknown constant. Our benchmark is
prior model (\ref{ssprior}), fitted by \texttt{varbvs} \cite[]{carbonetto2012scalable}, which requires to specify $q$.
We set $q=0.01$, implying that we expect a fairly sparse signal with a prior 99\% probability for $\beta_j$ to equal 0.
Then, to ensure a meaningful comparison with the benchmark model, we set $\bar{q} = p^{-1} \sum_{j=1}^p q_j = 0.01$, which determines $C$, resulting in absolute estimates of $q_j$. Fourth, these estimates are then used to define the feature-specific priors (\ref{ssprnew})and the spike-and-slab model is fit using the \texttt{varbvs} package.

\para
As a proof of concept, we show the benefit of moderating the prior inclusion probabilities by co-data in a high-dimensional SNP setting (minor allele frequencies (MAF), simulated as in \cite[]{carbonetto2012scalable}). For samples $i=1, \ldots, 500$ and features $j=1, \dots, p=10000$, generate:

\begin{alignat*}{2}
M_j &= 0.05 + 0.45U_j, U_j \sim^{\text{iid}} U[0,1] &\text{(MAFs)}\\
x_{ij} &\sim^{\text{iid}} \text{Bin}(2,M_j) &\text{(Allele counts)}\\
(\beta_j)_{j=1}^{150} &\sim^{\text{iid}} N(0,\tau^2=0.25); (\beta_j)_{j=151}^{p} = 0 &\text{(Coefficients)}\\
Y_i &= \sum_{j=1}^{10000} \beta_j x_{ij} &\text{(Response)}
\end{alignat*}

\noindent
Furthermore, the co-data is generated as follows. The grouping $Z_{1.}$ contains two groups: a small group of size 500 features of which 100 features correspond to
those with non-zero $\beta_j$'s, and 400 correspond to those with $\beta_j$'s equalling 0; and a large group of the remaining 9500 features, 50 of which correspond to non-zero $\beta_j$'s. Therefore, the first group is enriched in signal. The second co-data source $Z_{2.}$ consists of log (external) p-values. The first 150 p-values are generated from a balanced mixture of two beta distributions: $\mathcal{B}(0.1,10)$ and $\mathcal{B}(1,5)$, and the remaining ones where generated from a uniform distribution. Hence, this co-data source should also be informative, as the non-zero $\beta_j$'s tend to correspond to relatively small external p-values.

\begin{figure}[h]
\begin{center}
\includegraphics[scale=0.65]{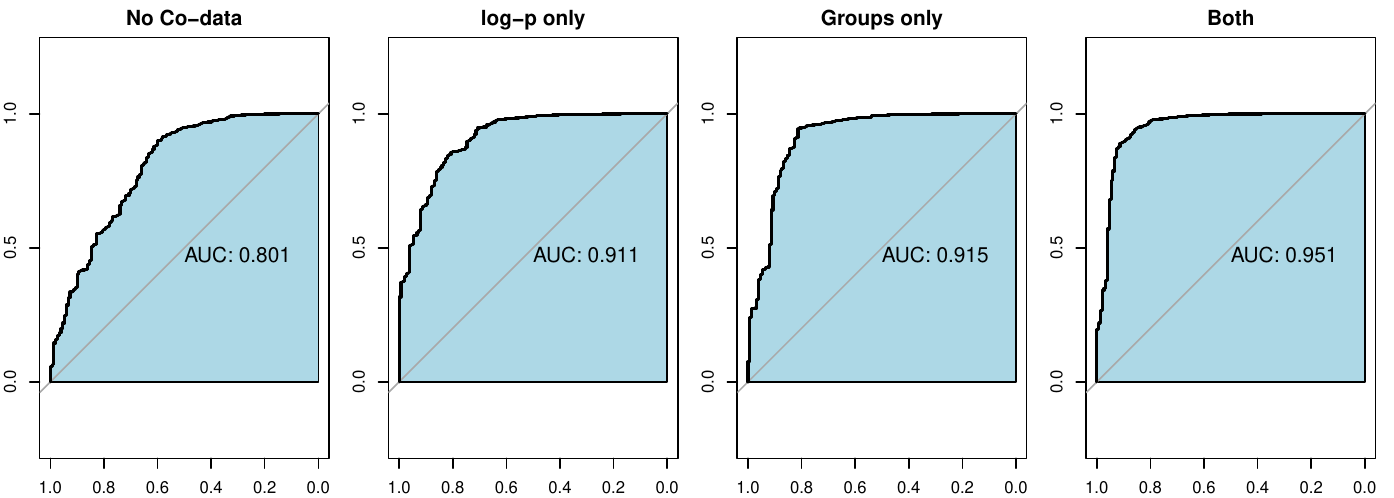}
\end{center}
\caption{ROC curves for feature selection using spike-and-slab models. X-axis: specificity, y-axis: sensitivity.}\label{sands}
\end{figure}

\para
For this simulated data set the co-data is indeed very informative. The estimated hyper-parameters $(\hat{\alpha}_1, \hat{\alpha}_2)$ render estimated prior inclusion probabilities $(\hat{q})_{j=1}^{150}$ with median: 0.136 and quartiles: (0.0096, 0.306) for relevant features. These are considerably higher than the summaries for the irrelevant ones, $(\hat{q})_{j=151}^{p}$, with median: 0.0031 and quartiles: (0.0028, 0.0078). Figure \ref{sands} shows the results for one simulated data set (as results are qualitatively very similar among multiple data sets). Not surprisingly, the improvement in feature selection performance with respect to the benchmark (no co-data) model is noticeable when either or both co-data sources are used.

\section{Software}
R-scripts to reproduce the results in this manuscript are available via: \url{https://github.com/markvdwiel/CodataReview/}.

%
%
%

\section{Discussion}
We reviewed methods that implement guided adaptive shrinkage, including group-adaptive methods. The latter framework was contrasted with group-regularized methods such as the sparse group-lasso. In genomics settings, multiple sources of co-data of mixed types, such as external p-values and known gene signatures, are often available. Fortunately, this setting is nowadays conveniently accommodated by several methods via a regression-type parametrization that links those co-data to the hyperparameters. We emphasize that the combination of \emph{multiple} co-data can be a very powerful tool to improve feature selection, as was illustrated for the simulated SNP data. Therefore, tools that accommodate automatic retrieval of co-data are a very welcome addition to the methodology. For this purpose, \cite{perscheid2021comprior} developed \texttt{Comprior}, which provides tools for extraction of gene and/or pathway scores or lists from several well-known genomic data bases. Moreover, it integrates with \texttt{xtune} to use such co-data for lasso-based feature selection.
In addition, \cite{wang2023prior} developed a module for automatic retrieval of gene-centered co-data from scientific articles. In short, they use a convolutional neural net based text analysis to learn a gene score for its relation to the disease of interest. This score may then be used as co-data for a given study.

\para
An important criticism on guided adaptive shrinkage methods is that they may be prone to overfitting when the number of hyperparameters (size of $\balpha$) is large. Full Bayesian methods like \texttt{graper} \cite[]{velten2018adaptive} and \texttt{infHS} \cite[]{busatto2023informative} may counter this by using a (weakly) informative prior, whereas \texttt{ecpc} \cite[]{van2023ecpc} allows to regularize the empirical Bayes moment equations to stabilize the estimation of $\balpha$. For the group-adaptive lasso setting we provide a potential solution in the Supplementary material, based on targeted hyperparameter shrinkage.

\para
We primarily focused on evaluating accuracy of feature selection, as most of the reviewed guided adaptive shrinkage methods contain extensive results on the potential
of these methods for improving prediction. Moreover, it has been demonstrated that also the stability of the selected feature set improves with the use of co-data \cite[]{van2023ecpc}.

\para
Finally, we note that the use of prior information is fundamental to science: `science builds on science'. The plethora of guided shrinkage methods reviewed here provides researchers the means to structurally do so in high-dimensional -omics settings.

\bibliographystyle{C://Synchr///Stylefiles/author_short3} 
\bibliography{C://Synchr//Bibfiles//bibarrays}      

\newpage

\section{Supplementary Material}
\section{Additional figures group-lasso vs group-adaptive lasso}
\begin{figure}[h]
\begin{center}
\includegraphics[scale=0.44]{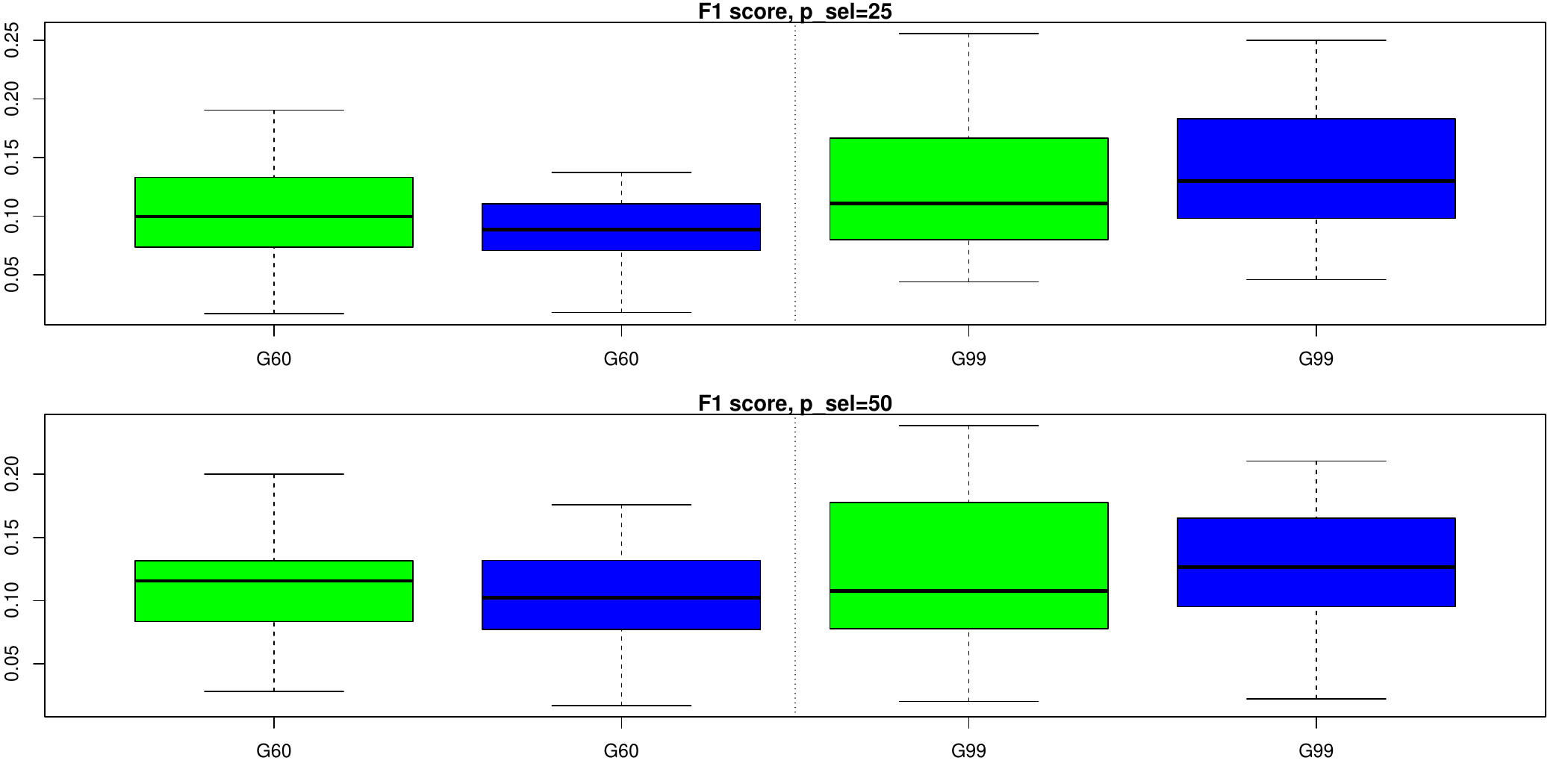}
\end{center}
\caption{F1 score for variable selection ($p_{\text{sel}} = 25, 50$) for group-adaptive lasso (green) and sparse group-lasso (blue) for $G =60, 99$ feature groups (25 simulations per $G$); group-sparse setting.}\label{F1sparse}
\end{figure}

Simulation settings. Figure \ref{F1sparse} corresponds to a group-sparse setting: $p=10,000, n = 200, G = 60, 99$; 5 groups contains signal.

\subsection{Targeted hyperparamer shrinkage for the group-adaptive lasso}
Below, we discuss an extension of the group-adaptive lasso  \cite[]{zeng2021incorporating, van2023fast} that target the shrinkage of the hyperparameters, the group-specific lasso penalties $(\lambda_g)_{g=1}^G$, by maximizing a penalized marginal likelihood based on a weakly informative prior.
Previously, we showed that the marginal likelihood with lasso priors can be approximated by use of central Gaussian priors which second moments matched to those of the double-exponential (=lasso) priors \cite[]{van2023fast}. Therefore, for stabilizing the group-adaptive lasso it suffices to perform the hyperparameter shrinkage on the level of the group-level \emph{ridge} penalties, which we denote by $(\lambda^R_g)_{g=1}^G$. For efficient optimization \texttt{squeezy} \cite[]{van2023fast} determines the log-marginal likelihood and its gradient, which are both additive in terms of the hyperparameters, with respect to $\lambda^R_{\text{log}} = \lambda^R_{\text{log},g} = \log \lambda^R_g$. We augment the log-marginal likelihood by a penalty, which is simply the sum of the log-priors of $\log \lambda^R_g$. For the latter, we use a Laplace prior with location $\log \lambda^R_\text{common}$ as target and scale 1. We opt for fixing the latter as this renders a very efficient algorithm that does not require further tuning, while still covering a wide range of potential values of $\lambda^R_g$. The Laplace prior is used to accommodate group-sparse settings. The target is obtained by applying \texttt{squeezy} for the $G=1$ setting, which is very fast.

As the Laplace prior has no gradient at its location parameter, we approximate the absolute value in this prior by $\sqrt(x^2+c)$ using a small value of $c$.
We experienced that this optimization is computationally very competitive to that of \texttt{squeezy}, which was reported to outperform other group-adaptive lasso methods, such as \texttt{gren} \cite[]{munch2021adaptive} and \texttt{ipf-lasso} \cite[]{boulesteix2017ipf}. In fact, we noticed that when the number of hyperparameters is large, the extra penalty helps to identify the optimum faster.

\begin{figure}[h]
\begin{center}
\includegraphics[scale=0.44]{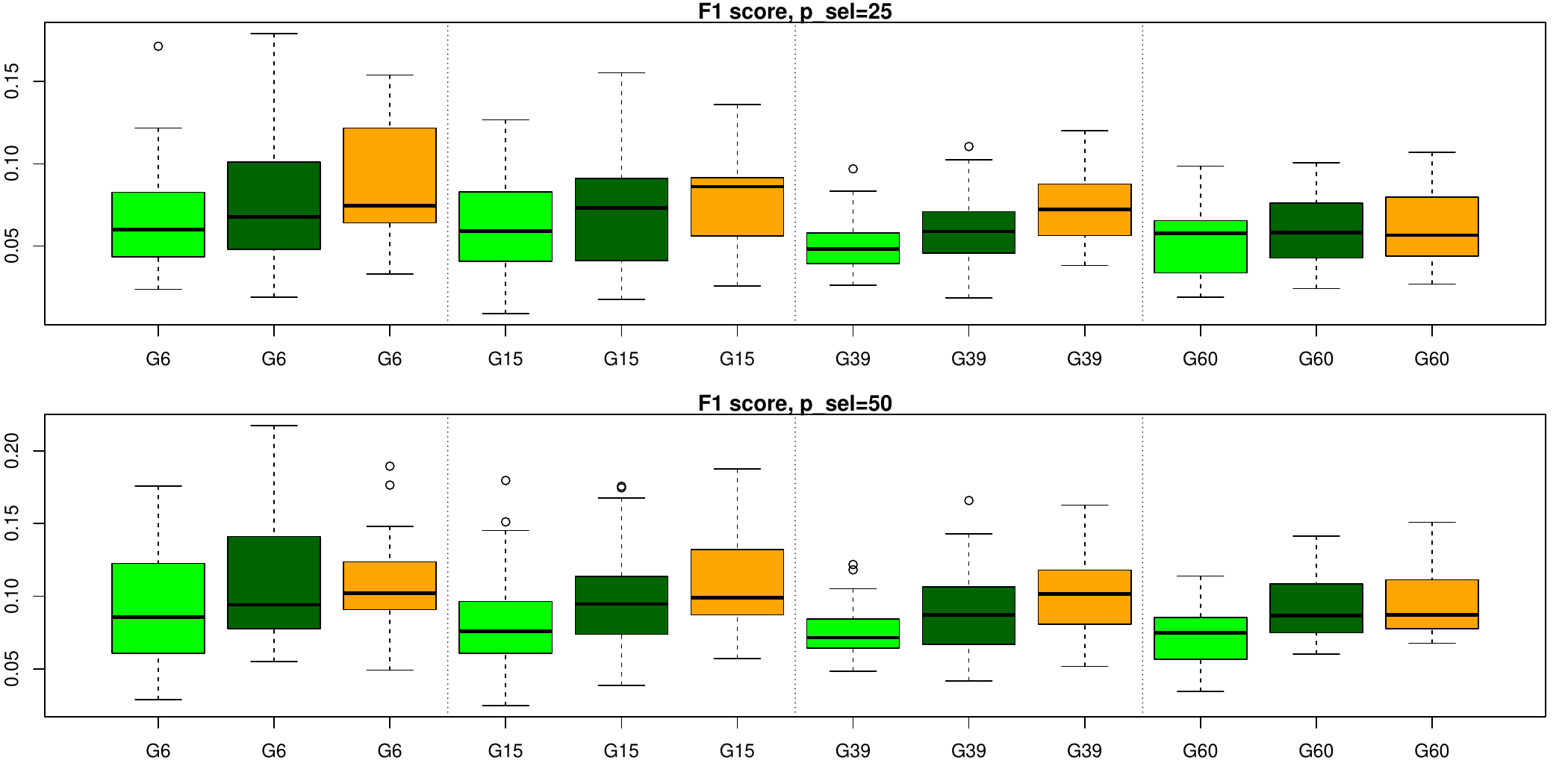}
\end{center}
\caption{F1 score for variable selection ($p_{\text{sel}} = 25, 50$) for group-adaptive lasso (green), targeted group-adaptive lasso (dark-green) and lasso (orange) for $G =6, 15, 39, 60$ \emph{non-informative} feature groups (25 simulations per $G$)}\label{F1R2random}
\end{figure}

\begin{figure}[h]
\begin{center}
\includegraphics[scale=0.44]{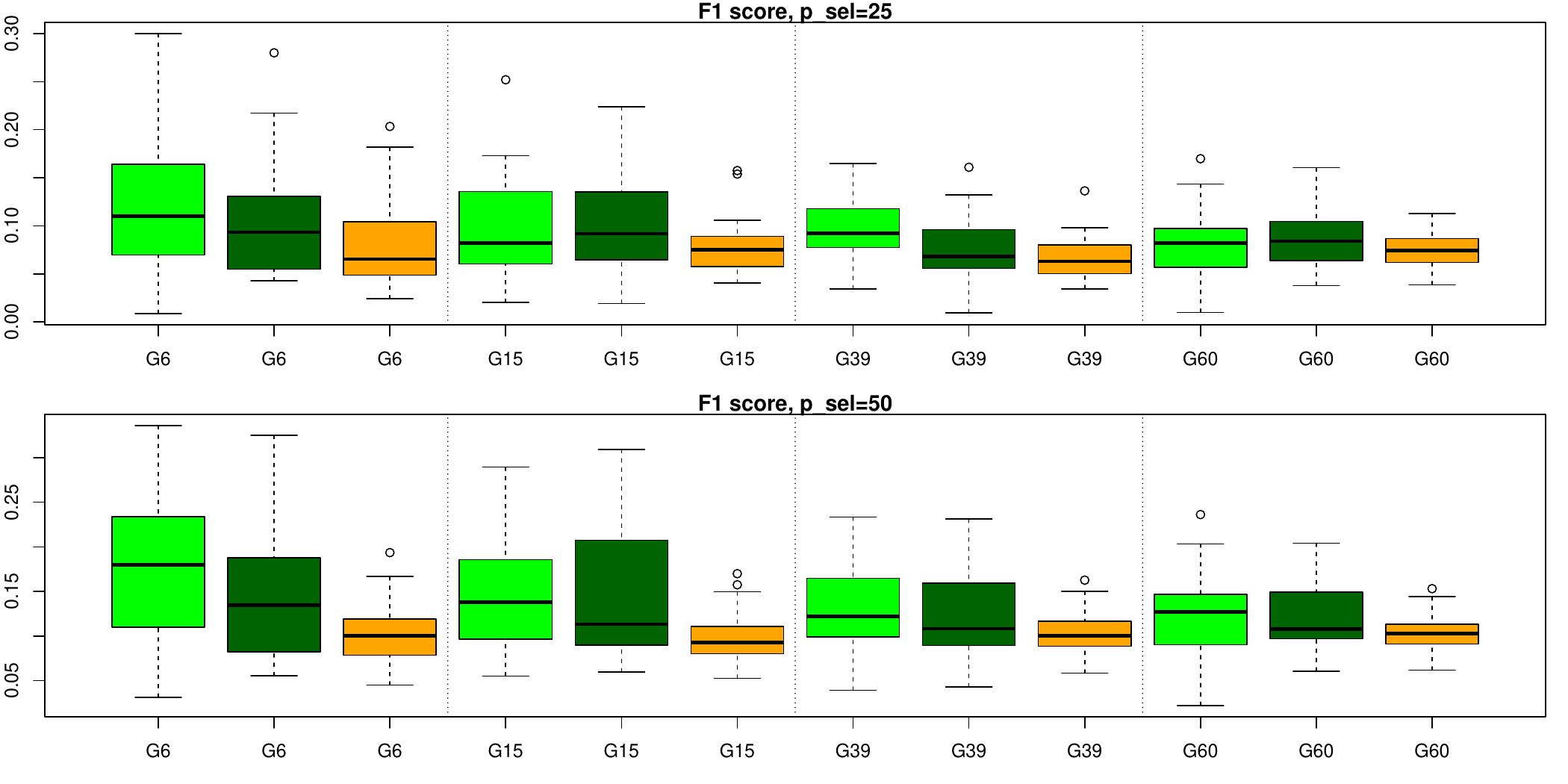}
\end{center}
\caption{F1 score for variable selection ($p_{\text{sel}} = 25, 50$) for group-adaptive lasso (green), targeted group-adaptive lasso (dark-green) and lasso (orange) for $G =6, 15, 39, 60$ informative feature groups (25 simulations per $G$)}\label{F1targeted}
\end{figure}

We implemented the approach by adapting \texttt{squeezy}. As over-fitting is particularly a concern when the grouping is not relevant at all, we first illustrate the results on this setting. To mimic such a setting we repeated the simulations as lined out in the Main Document for $G=6, 15, 39, 60$, but with non-zero features randomly assigned to all groups. Then, one would want the performance of a group-adaptive method to be close to that of its non-group-adaptive counterpart, here the ordinary lasso. Figure \ref{F1R2random} shows the results for feature selection. We observe that the targeted hyperparameter shrinkage improves results substantially in this `null setting', although results are still somewhat inferior those of the lasso, as the latter accommodates this setting best. In addition,  Figure \ref{F1targeted} shows that the targeted shrinkage of hyperparameters renders only minor loss of feature selection performance as compared to group-adaptive lasso without hyperparameter shrinkage when the groups are informative in the aforementioned simulation settings.

\end{document}